\documentstyle[aps,pre,epsf,epsfig]{revtex}
\begin{document}


\twocolumn[
\hsize\textwidth\columnwidth\hsize\csname @twocolumnfalse\endcsname

\title{
Grain boundary pinning and glassy dynamics in stripe phases
}
\author{
Denis Boyer
}
\address{ 
School of Computational Science and Information Technology,
Florida State University, Tallahassee, Florida 32306-4120.\\
{\rm (Present address:} Instituto de F\'\i sica, Universidad
Nacional Aut\'onoma de M\'exico, Apartado Postal 20-364, 
01000 M\'exico D.F., M\'exico {\rm )}
}
\author{Jorge Vi\~nals}
\address{Laboratory of Computational Genomics, Donald Danforth Plant Science
Center, 975 North Warson Rd, St. Louis, Missouri 63132.}
\date{\today}

\maketitle

\begin{abstract} 
We study numerically and analytically the coarsening of stripe phases 
in two spatial dimensions, and show that transient
configurations do not achieve long ranged orientational 
order but rather evolve into glassy configurations with 
very slow dynamics. 
In the absence of thermal fluctuations, defects such as
grain boundaries become pinned in an effective periodic potential that is
induced by the underlying periodicity of the stripe pattern itself. 
Pinning arises without quenched disorder from the non-adiabatic 
coupling between the slowly varying envelope of the order parameter 
around a defect, and its fast variation over the stripe wavelength.
The characteristic size of ordered domains asymptotes to a finite value 
$R_g \sim \lambda_0\ \epsilon^{-1/2}\exp(|a|/\sqrt{\epsilon})$, where
$\epsilon\ll 1$ is the dimensionless distance away from threshold,
$\lambda_0$ the stripe wavelength, and $a$ a constant of order unity.
Random fluctuations allow defect motion to resume until a new characteristic 
scale is reached, function of the intensity of the fluctuations. We finally
discuss the relationship between defect pinning and the coarsening
laws obtained in the intermediate time regime.
\end{abstract}

\narrowtext
]

\section{Introduction}

The motion of topological defects in two dimensional smectic phases is
studied at a finite distance from threshold. We focus on
the Swift-Hohenberg model of Rayleigh-B\'enard
convection and related amplitude equations to address the role that 
non-adiabatic effects 
play in domain coarsening of a modulated phase, defect pinning, and the 
appearance of glassy behavior.

Topological defects are often the longest lived modes of a non-equilibrium 
system, with their motion determining the longest relaxation times of the 
structure. Phenomenological models of defect motion that are based
on a mesoscopic description have been known
for some time \cite{re:kawasaki84a,re:brand84}. Such a description,
valid for distances much larger than the defect core, typically
involves time dependent Ginzburg-Landau equations or their 
generalizations. A few cases have been studied extensively, including domain 
coarsening in O(N) models \cite{re:mazenko95,re:bray95},
in nematics \cite{re:chuang93,re:toyoki93,re:zapotocky95,re:mason93},
and in smectic phases as effectively encountered in
models of Rayleigh-B\'enard convection or lamellar phases of block copolymers
\cite{re:oono88b,re:elder92,re:elder92b,re:cross95a,re:hou97,%
re:christensen98,re:boyer01b}. In the case of modulated phases,
the motion of a single defect has been
widely studied within the well known amplitude equation formalism.
This method describes the spatio temporal evolution of the envelope
of a base periodic or modulated structure
\cite{re:manneville90,re:cross93,re:siggia81b,re:manneville83b,re:tesauro87}.
The amplitude equation description is valid 
only close to bifurcation points where the spatial scale of variation 
of the amplitudes is large or \lq\lq slow" compared with 
the \lq\lq fast'' period of the base pattern and, in the present case, 
with the extent of the defect core as well.

Far enough from the bifurcation threshold of the modulated phase, 
the separation between slow and fast scales no longer holds, 
and corrections to the amplitude
equations appear because of the coupling between both scales. 
These corrections
are generically referred to as non-adiabatic effects.
One manifestation of non-adiabaticity is that a defect that would be
expected to move at constant velocity from an amplitude equation
analysis may instead remain immobile or pinned 
\cite{re:pomeau86,re:bensimon88b,re:malomed90}. We argue below that
non-adiabatic effects and defect pinning have important consequences 
for domain coarsening of modulated phases in two dimensions, 
and are responsible for the formation of glassy configurations.

Our results complement recent research on
glassy properties of stripe phases.
It has been suggested that systems in which long ranged order
is frustrated by repulsive interactions (the latter often leading to the
formation of stripe phases or other patterns in equilibrium) 
may in fact exhibit the properties of structural glasses. An example are
the glassy states recently observed in doped semiconductors 
in a stripe phase \cite{re:julien99}.
Coarse grained models with competing interactions of the type used here
(and also used to study block copolymer melts in lamellar phases) have
been reintroduced to describe
the formation of glasses in supercooled liquids \cite{re:kivelson95}.
Additional equilibrium studies of the same models in three dimensions 
based on replica calculations \cite{re:schmalian00} or Monte Carlo 
simulations \cite{re:grousson01} have been used to argue
for the existence of an equilibrium glass transition.
Structural glasses form spontaneously at low temperature without the
presence of any quenched disorder, and their properties remain in general
poorly understood. It is noteworthy that
coarse grained models exhibiting glassy behavior in the absence of
disorder are rare, whereas examples of discrete systems are known
({\it e.g.} Ising models with next-nearest neighbor interactions
\cite{re:shore92,re:newman99}).
We present here a {\it two} dimensional study that indicates a
dynamical route to the formation of glassy configurations in 
stripe phases.

We first analyze the motion of a particular type of defect, namely a
grain boundary separating two domains of differently oriented stripes.
Earlier asymptotic work near onset ({\it i.e.} in the 
limit $\epsilon \rightarrow 0$, where $\epsilon$ is the dimensionless 
distance away from threshold) is extended to the region of small but 
finite $\epsilon$. In Section \ref{sec:pinning}, grain boundaries are 
shown to move in an effective 
periodic potential of wavelength $\lambda_0/2$ (where $\lambda_0$ is 
the periodicity of the stripe modulation) and of magnitude that
increases very quickly with $\epsilon$.
Grain boundaries asymptotically pin as the 
driving force for grain boundary motion decreases.  
It is argued that
for any finite $\epsilon$ an infinite size system will not achieve 
macroscopic long range order dynamically following a quench. 
Rather, the characteristic size of
a domain will not exceed typical value $R_g$ that is proportional to
$\lambda_0\ \epsilon^{-1/2}\exp(|a|/\sqrt{\epsilon})$, where $a$ is
a constant of order unity.

In Section \ref{sec:finite_temp}, we incorporate the effect of random
fluctuations and derive the corresponding
amplitude equations valid for fluctuations of small amplitude.
The asymptotic motion of a grain boundary can be recast as an
escape problem in which the effective activation barrier
is seen to be proportional to the grain boundary perimeter.

Our approach must be considered only qualitative in
nature because of the scope of the description employed. Ginzburg-Landau
equations, and more generally amplitude or order parameter equations
(of which the Swift-Hohenberg model described below is but one
example) are only asymptotic, large length scale approximations to the
physical system
they model in the immediate vicinity of a bifurcation point.
Therefore any short scale phenomena involved
in the description of non-adiabatic corrections clearly falls beyond
their range of validity, at least in a systematically quantifiable
way. It is nevertheless not unreasonable to expect that non-adiabatic
effects of the sort encountered in order parameter
equations will also occur in the physical systems
which they model. Furthermore, our results also provide insights
into many existing numerical studies of these order parameter models, as
described below.

In Section \ref{sec:coarsening}, we address the consequences of pinning 
on the domain coarsening that occurs in the intermediate time regime 
following the quench.
This subject has been the focus of several numerical studies
\cite{re:elder92,re:elder92b,re:cross95a,re:hou97,re:christensen98,re:boyer01b}
and, more recently, of experimental studies in block copolymer thin films
\cite{re:harrison00b} and in electro-convection in nematics \cite{re:purvis01}.
The results of Sections \ref{sec:pinning} and
\ref{sec:finite_temp} provide a possible interpretation of conflicting
results in the literature. Previous studies of this problem
\cite{re:elder92,re:elder92b,re:cross95a,re:hou97,re:christensen98,re:boyer01b}
addressed the existence of self-similarity during domain coarsening and
attempted to quantify the time dependence of the linear scale of the
coarsening structure. The statistical self-similarity hypothesis asserts
that after a possible transient, consecutive configurations of the
coarsening structure are geometrically similar in a statistical sense.
As a consequence, any linear scale of the structure (e.g., the average
size of a domain or grain of like oriented stripes) is expected to grow
as a power law of time $l(t) \sim t^{1/z}$, with $z$ a characteristic
exponent. Self-similarity is a well known feature in systems that order
in uniform phases of broken symmetry \cite{re:gunton83,re:bray94}.
However, the determination of $z$ has been problematic for
stripe phases. Its value appears to depend on the quench depth (the value of
$\epsilon$), whether or not fluctuations are included in the
governing equations, on the thermal history of the system, and on 
the particular linear scale analyzed. 

Recent work in the limit $\epsilon \rightarrow 0$ showed that
coarsening is self-similar and that $z = 3$ \cite{re:boyer01b}.
The value $z=3$ in that limit can
be justified by a dimensional analysis of the
law of grain boundary motion. We focus here on the case of
finite $\epsilon$ (in practice
$\epsilon \ge 0.1$ for the Swift-Hohenberg model),
and report a slowing down of phase ordering dynamics with increasing
$\epsilon$, in agreement with the literature.
We attribute this behavior to partial pinning of defects that
becomes increasingly important at long times as the driving force for
coarsening decreases. At even longer times, coarsening stops altogether
and the system reaches a glassy state as the linear scale of the
structure reaches the critical value $R_{g}(\epsilon)$ computed in 
Section \ref{sec:pinning}.  
When random fluctuations are incorporated in the model, we show that, 
sufficiently close to onset, the value of $z$ remains independent of the 
intensity of the fluctuations, thus verifying the universality implied in 
the self-similarity hypothesis in that region.
At larger $\epsilon$, we find that fluctuations accelerate
ordering kinetics, also in agreement with the literature, and that, as 
expected, defect motion is allowed beyond the scale given by $R_{g}$. 
At even later times the system orders very slowly, possibly
logarithmically in time.
 
\section{Non-adiabatic corrections and grain boundary pinning}
\label{sec:pinning}

We consider the Swift-Hohenberg model
of Rayleigh-B\'enard convection \cite{re:swift77} as a
prototypical model of a modulated phase.
The numerical results presented below have been obtained
from a direct numerical solution of the model. The analytic
results, on the other hand, follow from the corresponding
amplitude equation, and hence are expected to be of somewhat
wider generality. The model equation studied here is
\begin{equation}
\label{sh}
\frac{\partial \psi}{\partial t}=
\epsilon\psi-\frac{1}{k_0^4} (k_0^2+\nabla^2)^2\psi-\psi^3\ ,
\end{equation}
where $\psi$ is an dimensionless order parameter related to the vertical
fluid velocity at the mid plane of a Rayleigh-B\'enard convection cell,
$\epsilon$ is the reduced Rayleigh number $(R-R_c)/R_c \ll 1$
($R_{c}$ is the critical Rayleigh number for instability),
and $k_{0}=2\pi/\lambda_0$ is the roll wavenumber (in Appendix A we
outline the connection between
this model and other coarse-grained models with
long range repulsive interactions \cite{re:leibler80}).

\begin{figure}  
\epsfig{figure=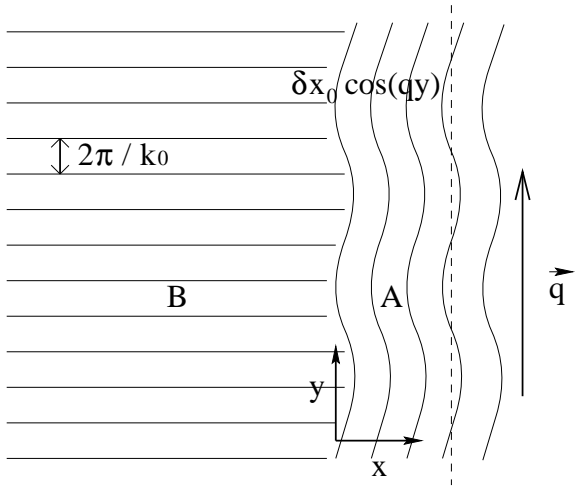,width=3in}
\vspace{1.0cm}
\caption{Schematic grain boundary configuration separating two domains of 
stripes $A$ and $B$ of the same periodicity 
($|\vec{k}_0|=|\vec{k}_{0}^{\ \prime}|=k_0$). 
The stripes of domain $A$ are weakly curved by a transverse modulation 
of wavenumber $q \ll k_0$. $\delta x_0$ represents the magnitude of the
phase modulation.}
\label{figgb}
\end{figure}

\begin{figure}
\epsfig{figure=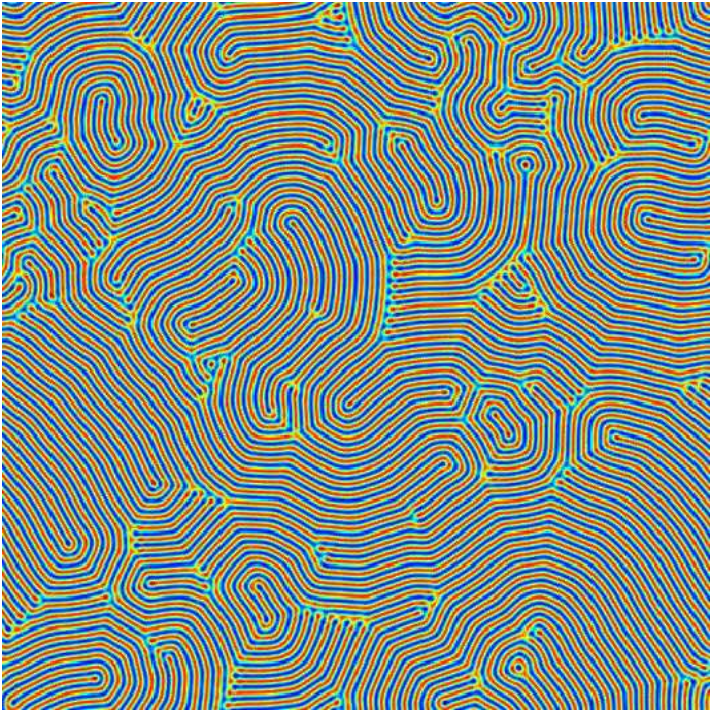,width=3.7in}
\vskip 0.5cm
\epsfig{figure=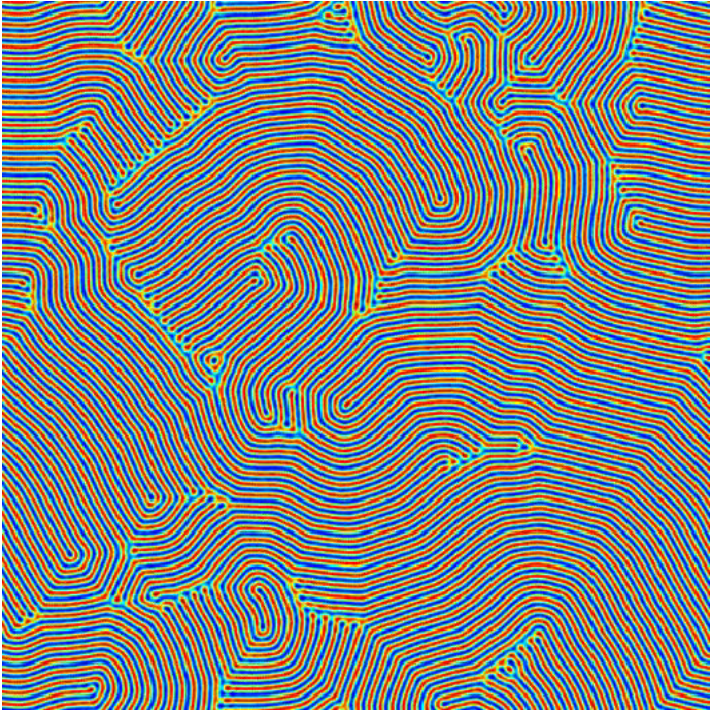,width=3.7in}
\vspace{1.0cm}
\caption{(a) Near stationary configuration obtained
after a quench at $\epsilon=0.4$ and in the absence of
fluctuations $F=0$ (the time shown is $t=2.3\times 10^5$, and the system size
includes $512^2$ nodes). (b) New structure obtained after taking the
configuration shown in (a) as an initial condition and further integrating the
model equations with $F=0.00318$ for a period of $10^5$ time units. At this
time, any boundary motion is very slow.}
\label{figheating}
\end{figure}

For $0 < \epsilon \ll 1$, the leading order approximation to the
stationary solution of Eq. (\ref{sh}) is a sinusoidal function of wavenumber
$k_{0}$.
We focus in this section on a configuration that contains an isolated
grain boundary separating two
such stationary solutions with mutually perpendicular wavevectors
(Figure \ref{figgb}). The reason for studying this perpendicular orientation
is the expectation that a $90^{\circ}$ grain boundary is that
of lowest energy, and hence the prevalent boundary angle in
an extended system that
evolves spontaneously from an initially disordered configuration
(see, for example, Figure \ref{figheating}a,b). It is known that a planar
grain boundary separating two regions of uniform $k_{0}$ is stationary
\cite{re:manneville83b,re:tesauro87}. However,
we found in ref. \cite{re:boyer01} that a slightly perturbed
boundary undergoes a net translation with a speed that is a function of the
curvature of the rolls ahead of it. We address in this section the extension
of the asymptotic results given in that reference to small but finite
$\epsilon$, and show how corrections obtained lead to
boundary pinning.

Near threshold, a $90^{\circ}$ grain boundary configuration is an
approximate solution of Eq. (\ref{sh}) of the form
\begin{eqnarray}
\label{psi}
\psi(x,y,t)&=&\frac{1}{2}\left[ A(X_A,Y_A,T)\ e^{ik_0x} \right.\nonumber\\
&+& \left.B(X_B,Y_B,T)\ e^{ik_0y}+ c.c\right],
\end{eqnarray}
where slow variables are denoted by capital letters and are defined as
\cite{re:manneville83b,re:tesauro87}
\begin{eqnarray}\label{slow}
X_A&=&\epsilon^{1/2}x,\ Y_A=\epsilon^{1/4}y;\nonumber\\ 
X_B&=&\epsilon^{1/4}x,\ Y_B=\epsilon^{1/2}y;\ T=\epsilon t.
\end{eqnarray}
(The coordinate $x$ is directed along the normal to the reference planar
boundary.)

We recall first some known results for a planar and
stationary grain boundary in the limit $\epsilon \rightarrow 0$, a case that
was extensively studied in refs. \cite{re:manneville83b,re:tesauro87}.
The stationary amplitudes $\{A_0,B_0\}$ are a function only of $x$.
$A_0$, the amplitude of the
rolls parallel to the interface, vanishes as $\exp(x\sqrt{\epsilon}/\lambda_0)$
when $x \rightarrow -\infty$, and saturates to $(4\epsilon/3)^{1/2}
{\rm tanh}(x\sqrt{\epsilon}/\lambda_0)$ when $x\rightarrow+\infty$.
The behavior of the amplitude of the rolls perpendicular to the interface
is slightly different: $B_0(x)-(4\epsilon/3)^{1/2}\propto
\exp(x\sqrt{\epsilon}/\xi_0)$ when $x\rightarrow-\infty$ and there exists
a location $x^*$ such that $B_0(x>x^*) \simeq 0$ to a good approximation.
Hence, the grain boundary region has a thickness proportional
to $\lambda_0/\sqrt{\epsilon}$. It is important to note that at small
$\epsilon$ the location of the grain boundary decouples from the
phase of the stripes of domain $A$. Thus, the configuration obtained
is invariant under any translation of the grain boundary by a distance $x_0$
(the phase of the stripes remaining unchanged).

We next derive two coupled equations for the amplitudes $A$ and $B$ that
take into account the possible coupling between
these amplitudes and the phases of the stripes. This coupling becomes
significant at a finite value of $\epsilon$, and hence when there is
a large but finite separation
between the scales $\{X_{A,B},Y_{A,B}\}$ and $\{x,y\}$.
We follow an approach similar to that used in ref. \cite{re:malomed90} to
study the
motion of a planar front between a hexagonal and a uniform phase,
or between a hexagonal and a stripe phase.
The first step is a multiscale analysis, and is standard
\cite{re:manneville90}.
Equation (\ref{sh}) is expanded in power series of $\epsilon$, as well as
the solution $\psi=\epsilon^{1/2}\psi_{1/2}+\epsilon\psi_1+
\epsilon^{3/2}\psi_{3/2}+...$. The leading order solution
$\epsilon^{1/2}\psi_{1/2}$
is given by Eq. (\ref{psi}). At order $\epsilon^{3/2}$,
the solvability conditions for the existence of a nontrivial
solution for $\psi_{3/2}$
yield the relations that $A$ and $B$ must satisfy,
\begin{eqnarray}
\frac{1}{\lambda_0^2}\int_x^{x+\lambda_0} dx^{\prime}
\int_y^{y+\lambda_0} dy^{\prime}
\left[{\cal L}(\psi_{1/2})-\psi_{1/2}^3\right] e^{-ik_0x^{\prime}}=0\label{sca}\\
\frac{1}{\lambda_0^2}\int_x^{x+\lambda_0} dx^{\prime}
\int_y^{y+\lambda_0} dy^{\prime}
\left[{\cal L}(\psi_{1/2})-\psi_{1/2}^3\right] e^{-ik_0y^{\prime}}=0\label{scb}
\end{eqnarray}
with the linear operator ${\cal L}=1-\partial_T-k_0^{-4}
(\partial_{X_B}^2+\partial_{Y_A}^2+2\partial_y\partial_{Y_B}
+2\partial_x\partial_{X_A})^2 $. In the limit $\epsilon \rightarrow 0$ the
functions $A$ and $B$ remain constant over one spatial period $\lambda_0$,
and therefore the only non vanishing contribution to the integrals come
from the terms proportional to $e^{ik_0x^{\prime}}$ (resp.
$e^{ik_0y^{\prime}}$) within brackets in Eq. (\ref{sca})
(resp. Eq. (\ref{scb})). This standard
set of coupled Ginzburg-Landau
equations follows \cite{re:manneville83b,re:manneville90} .
It is known, however, that additional non perturbative
contributions arising from the term $\psi_{1/2}^3$ appear in
Eqs. (\ref{sca}) and (\ref{scb}). We focus next on these contribution and their
effect on the relaxation of a slightly perturbed grain boundary.

Integrals of the type
$\int_x^{x+\lambda_0} dx^{\prime}\ e^{imk_0x{\prime}} A^nB^p$
in Eqs. (\ref{sca}) and (\ref{scb})
(where $m$, $n$ and $p$ are integers) will not integrate to zero if
the thickness of the grain boundary profiles along the
$x$ direction is finite. 
(Contributions from the direction transverse to the grain boundary, 
$\int_y^{y+\lambda_0} dy^{\prime}\ e^{imk_0y^{\prime}} A^nB^p$,
will be neglected. They are typically of the order of $B^2\partial_y^2A$ and
hence always smaller than the leading analytical terms of the amplitude 
equations.) Terms proportional
to $e^{imk_0x^{\prime}}$ will contribute to Eq. (\ref{sca}), and
terms proportional to $e^{imk_0x^{\prime}+ik_{0}y^{\prime}}$ to Eq.
(\ref{scb}).
If we only retain the lowest order term $\epsilon^{1/2}\psi_{1/2}$ as
given by Eq. (\ref{psi}), we find that only $A^3$ ($m=3$) contributes
to Eq. (\ref{sca}), while $3A^2B$ ($m=2$),
as well as $3\bar{A}^2B$ ($m=-2$, with $\bar{A}$ the complex conjugate of $A$),
to Eq. (\ref{scb}). Reintroducing the original unscaled variables,
the generalized amplitude equations read
\begin{eqnarray}
\label{ampa}
\frac{\partial A}{\partial t}=&-&\frac{\delta F_{gb}}{\delta \bar{A}}\nonumber\\
&-&\frac{1}{4\lambda_0^2} \int_{x}^{x+\lambda_0}{dx'}
\int_{y}^{y+\lambda_0}{dy'}\ A^3(x',y',t)\ e^{i2k_0x'} ,
\end{eqnarray}
\begin{eqnarray}
\label{ampb}
\frac{\partial B}{\partial t}=&-&\frac{\delta F_{gb}}{\delta \bar{B}}\nonumber\\
&-&\frac{3}{4\lambda_0^2} \int_{x}^{x+\lambda_0}{dx'}
\int_{y}^{y+\lambda_0}{dy'}[A^2B e^{i2k_0x'}\nonumber\\ 
&+&\bar{A}^2B e^{-i2k_0x'}]\ ,
\end{eqnarray}
where $ F_{gb} = \int d \vec{r} {\cal F}_{gb}$
is the standard Lyapunov functional corresponding to the $90^{\circ}$ grain
boundary.
Its variational derivatives satisfy \cite{re:manneville83b,re:tesauro87}
\begin{eqnarray}
-\delta F_{gb}/\delta \bar{A}=&\epsilon& A+\frac{4}{k_0^2}
\left(\partial_x-\frac{i}{2k_0}\partial_y^2\right)^2A\nonumber\\
&-&\frac{3}{4}|A|^2A-\frac{3}{2}|B|^2A\ , \label{liapa}\\
-\delta F_{gb}/\delta \bar{B}=&\epsilon& B+\frac{4}{k_0^2}
\left(\partial_y-\frac{i}{2k_0}\partial_x^2\right)^2B\nonumber\\
&-&\frac{3}{4}|B|^2B-\frac{3}{2}|A|^2B\ .\label{liapb}
\end{eqnarray}
The last terms in the right hand sides of Eqs. (\ref{ampa}) and (\ref{ampb})
depend on both fast and slow spatial scales, and they embody the so-called
non-adiabatic coupling between the two. Analyzing the effects of these
two terms on the relaxation of a perturbed grain boundary is the subject of
remainder of this section.

We now introduce a small perturbation to the planar boundary as shown
schematically in Figure \ref{figgb}.
The phase of the stripes of domain $A$
is distorted by a uniform perturbation of wavenumber $q \ll k_0$ (and of
amplitude
$\delta x_0\ll\lambda_0$) in the direction transverse to the stripes.
As shown in ref.
\cite{re:boyer01}, approximate solutions to Eq. (\ref{ampa})-(\ref{ampb})
are given by
\begin{eqnarray}
A&=&A_0(x-x_{gb}(t))\ e^{ik_0\delta x_0\cos(qy)}\ ,
\label{sola} \\
B&=&B_0(x-x_{gb}(t))\ ,
\label{solb}
\end{eqnarray}
where $x_{gb}(t)$ represents the time dependent position of the grain
boundary (averaged over $y$). As already discussed in that reference,
perturbations to the phase of B are of higher order in
$\epsilon$. In order to derive a law of motion for
$x_{gb}$ it is simpler to neglect the linear relaxation of the perturbed rolls,
and hence assume that $\delta x_0$ is constant. The amplitude
$\delta x_{0}$ relaxes exponentially with time but the relaxation time of
the perturbation is proportional to $q^{-4}$ and usually much longer than the
characteristic time associated with grain boundary motion,
$\lambda_0/\dot{x}_{gb}$. Furthermore, as was shown in
ref. \cite{re:boyer01}, explicitly considering stripe relaxation
does not change the law of motion for $x_{gb}$ in any quantitative way.

Multiply Eq. (\ref{ampa}) (resp. Eq. (\ref{ampb})) by
$\partial_t \bar{A}$  (resp. $\partial_t \bar{B}$), add the results and
integrate the real part over the system area. By using Eqs.
(\ref{sola})-(\ref{solb}) and integrating by parts the non-adiabatic
terms, we obtain the following law of motion for the grain boundary
\begin{equation}
\label{motionlaw}
\dot{x}_{gb}=\frac{\epsilon}{3k_0^2D(\epsilon)}\ \kappa^2\
- \frac{p(\epsilon)}{D(\epsilon)}\cos(2k_0 x_{gb}+\phi)\ ,
\end{equation}
where $\kappa = \delta x_0 q^2$ is proportional to the mean
curvature of the stripes of domain $A$, $\phi$ is a constant phase, and
\begin{eqnarray}
&D&(\epsilon)= \int_{-\infty}^{\infty}
dx\ [(\partial_x A_0)^2+(\partial_x B_0)^2]\ , \label{D}\\
&p&(\epsilon)={\rm Max}_{\theta}\left\{
\frac{3}{4}\int_{-\infty}^{\infty} dx\ A_0^3(x)
\partial_xA_0(x)\cos(2k_0x+\theta)\right. \nonumber\\
&+&\left.
\frac{3}{2}\int_{-\infty}^{\infty} dx\ [2A_0B_0^2\partial_xA_0+A_0^2B_0
\partial_xB_0]\cos(2k_0x+\theta)\right\}.
\label{p}
\end{eqnarray}
Equation (\ref{motionlaw}) without the oscillatory term
was derived in Ref. \cite{re:boyer01} in the limit $\epsilon \rightarrow 0$.
The coefficient $D(\epsilon)$, with dimensions of an inverse length,
represents a friction term that depends
on the static grain boundary profile $\{A_0,B_0\}$, while the term
$\epsilon\kappa^2$ in the numerator is proportional to
$\int_0^L dy\ [{\cal F}_{gb}(x=\infty,y)-{\cal F}_{gb}(x=-\infty,y)]/L$, where
${\cal F}_{gb}$ is the free energy density implicitly defined by
Eqs. (\ref{liapa})-(\ref{liapb})).
The numerator can be understood as the leading contribution
(in $\epsilon$ and $\kappa$) from an external force acting on the grain
boundary. This force results from the difference in the free energy
density ${\cal F}_{gb}$ between curved stripes on one side, and straight
stripes on the other side of the boundary. Note the unusual
dependence of $\dot{x}_{gb}$ on a even power of the curvature thus 
indicating that the motion of the
grain boundary is such that curved parallel rolls of higher energy are
always replaced by straight perpendicular rolls.

The last term in the right hand side of Eq. (\ref{motionlaw}) is the
dominant contribution arising from the non-adiabatic terms of
Eqs. (\ref{ampa})-(\ref{ampb}). The {\it dimensionless} quantity
$p(\epsilon)$ plays the role of the amplitude of a periodic
potential of period $\lambda_0/2$ within which the grain boundary moves.
The major contribution to $p(\epsilon)$ comes
from the integral that contains the term $\partial_xB_0$ in Eq. (\ref{p}) since
the profile $B_0(x)$ has a steeper variation than $A_0(x)$
\cite{re:manneville83b}.
Given that both amplitudes $A_0$ and $B_0$ are approximately of the form
$\sqrt{\epsilon} f(\sqrt{\epsilon}x/\lambda_0)$,
it is easy to show from Eq. (\ref{p}) that
\begin{equation}
\label{estimp}
p(\epsilon)\ \sim \ \epsilon^2\ e^{-|\alpha|/\sqrt{\epsilon}}\ ,
\end{equation}
where $|\alpha|$ is a constant of order unity, corresponding to
the pole of the envelopes closest to the real axis in the complex plane.
Hence, $p$ behaves non-analytically at small $\epsilon$,
and increases extremely quickly with $\epsilon$.
Qualitatively similar results were reported in
\cite{re:bensimon88b,re:malomed90} for one dimensional fronts between
conductive and convective states, or between different convective states.

Equation (\ref{motionlaw}) shows that
for any finite $\epsilon>0$ a planar grain boundary ($\kappa=0$) can have only
two stationary positions per period of the stripe pattern $\lambda_0$.
This effect had been observed numerically and reported in ref.
\cite{re:boyer01}, with similar findings also given in ref.
\cite{re:tesauro87}.
Equation (\ref{motionlaw}) also implies that there exists a critical curvature
$\kappa_g$ below which the grain boundary will remain immobile.
This critical curvature is given by,
\begin{equation}
\label{rg}
\kappa_g = \frac{1}{R_{g}}=k_0\left(\frac{3p(\epsilon)}
{\epsilon}\right)^{1/2} \ ,
\end{equation}
where $R_{g}$ is the associated radius of curvature which diverges
non-analytically near onset
\begin{equation}
\label{estimrg}
R_g\sim\ \lambda_0\ \epsilon^{-1/2}\
\exp\left(\frac{|\alpha|}{2\sqrt{\epsilon}} \right)\ .
\end{equation}

\begin{figure}
\epsfig{figure=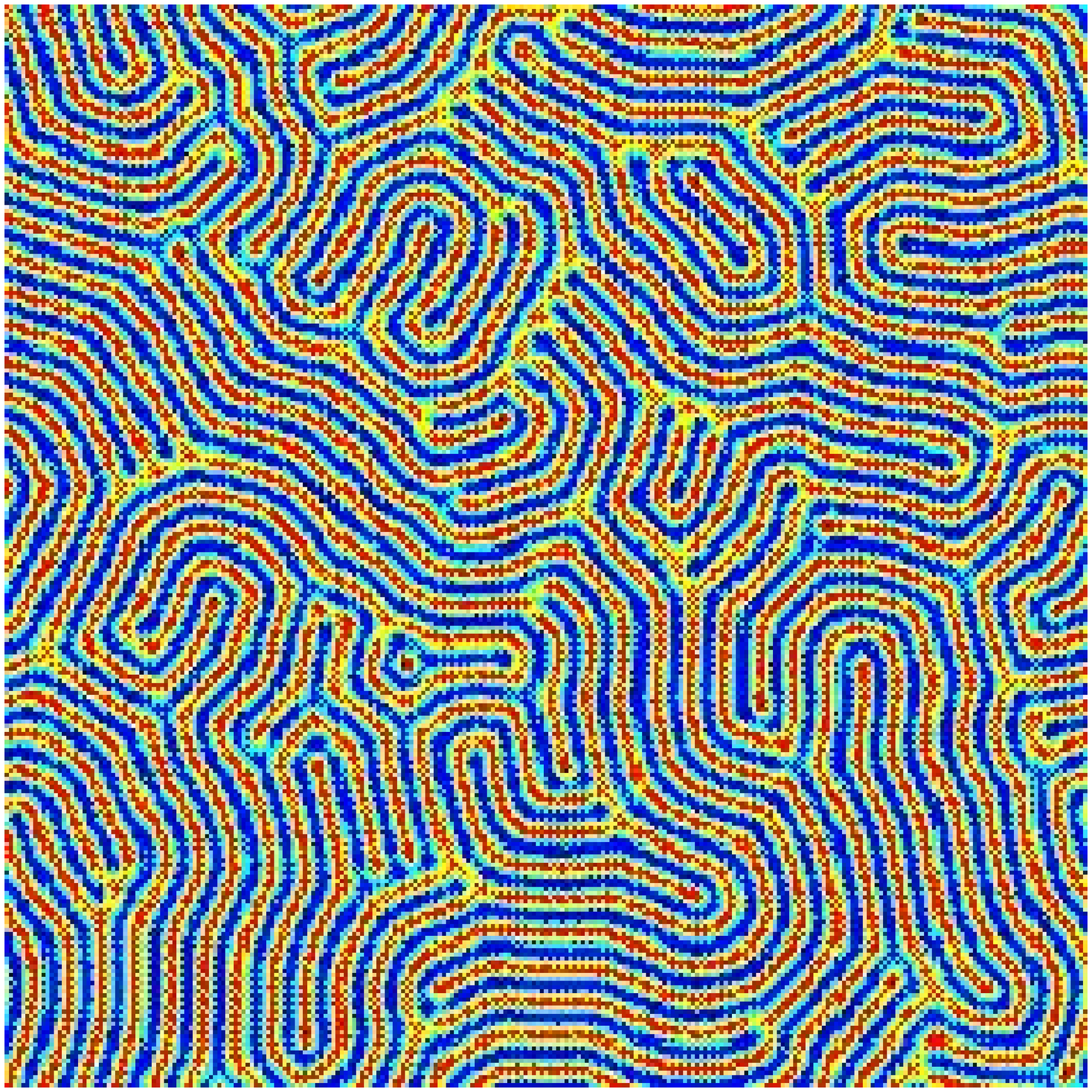,width=3in}\hskip0.5cm \epsfig{figure=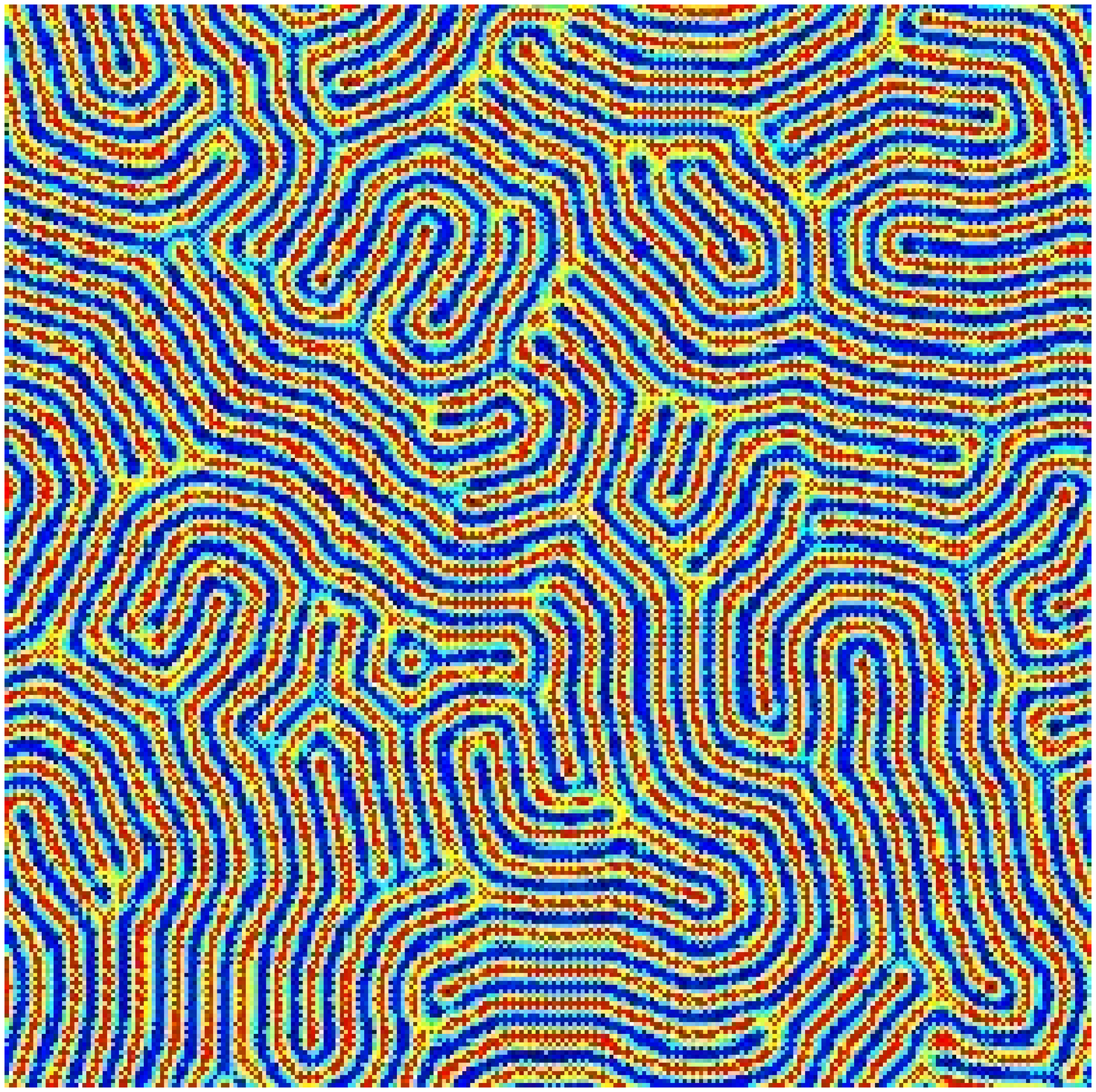,width=3in}
\vspace{1.0cm}
\caption{Glassy configurations obtained by numerical solution of the 
Swift-Hohenberg model with random initial conditions. Dimensionless
times shown are (a), $t=10000$; and (b), $t=20000$. 
Here $\epsilon=0.5$ and the system has $256^2$ grid nodes.}
\label{fignumRg}
\end{figure}

These results have been verified by direct numerical solution
of the Swift-Hohenberg model with reasonably small values of $\epsilon$.
The numerical algorithm used has been described
in \cite{re:boyer01,re:boyer01b}. Briefly,
Eq. (\ref{sh}) is discretized on a square grid of mesh size
$\Delta x = 1$ with $512^2$ nodes ($256^2$ for $\epsilon=0.5$), and
the wavelength is set to $\lambda_0 = 8 \Delta x$. A semi-implicit
spectral method is used to iterate in time.
The initial condition for $\psi$ is a white and Gaussian
random field with zero average and variance
$\langle \psi^{2} \rangle =\epsilon$.
Typical long time configurations which are stationary for all
practical purposes are shown in
Figs. \ref{fignumRg} and \ref{figheating}a. These figures show
the field $\psi$ in grey scale.
Many topological defects including dislocations, +1/2 disclinations and
several $90^{\circ}$ grain boundaries can be identified.
Figure \ref{fignumRg} corresponds to
$\epsilon=0.5$ and two different times $t=10^4$ and $t=2\times 10^4$,
showing that
the order parameter does not change beyond $t=10^4$.
Figure \ref{figheating}a corresponds to $\epsilon=0.4$, and
the configuration shown remains practically constant beyond
$t = 2.3\times 10^5$.

\begin{figure}
\epsfig{figure=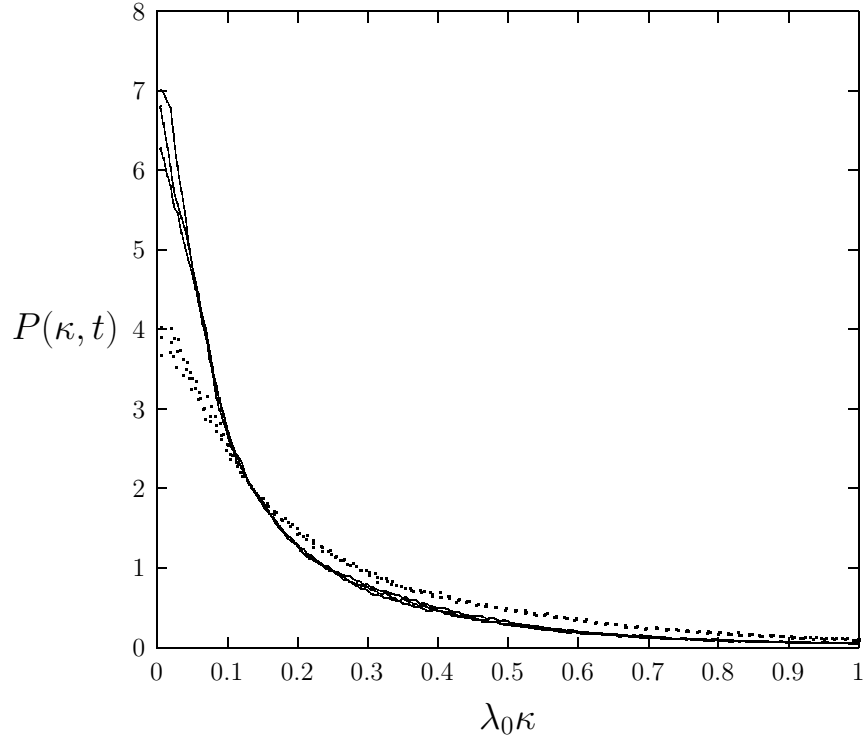,width=7in}
\vspace{-4cm}
\caption{Probability distribution function of stripe curvatures, 
$P(\kappa,t)$, 
after a quench at $\epsilon=0.5$ (dotted lines) and $\epsilon=0.4$
(solid lines), averaged over 10 and 6 independent runs respectively. 
For $\epsilon=0.5$ the figure shows the curves obtained at times
$t=10^4$, $5 \times 10^4$ and $10^5$, and for $\epsilon=0.4$ 
at times $t=6\times 10^4$, $1.2\times 10^5$ and $2.3\times 10^5$.}
\label{figcurv}
\end{figure}

\begin{figure}
\epsfig{figure=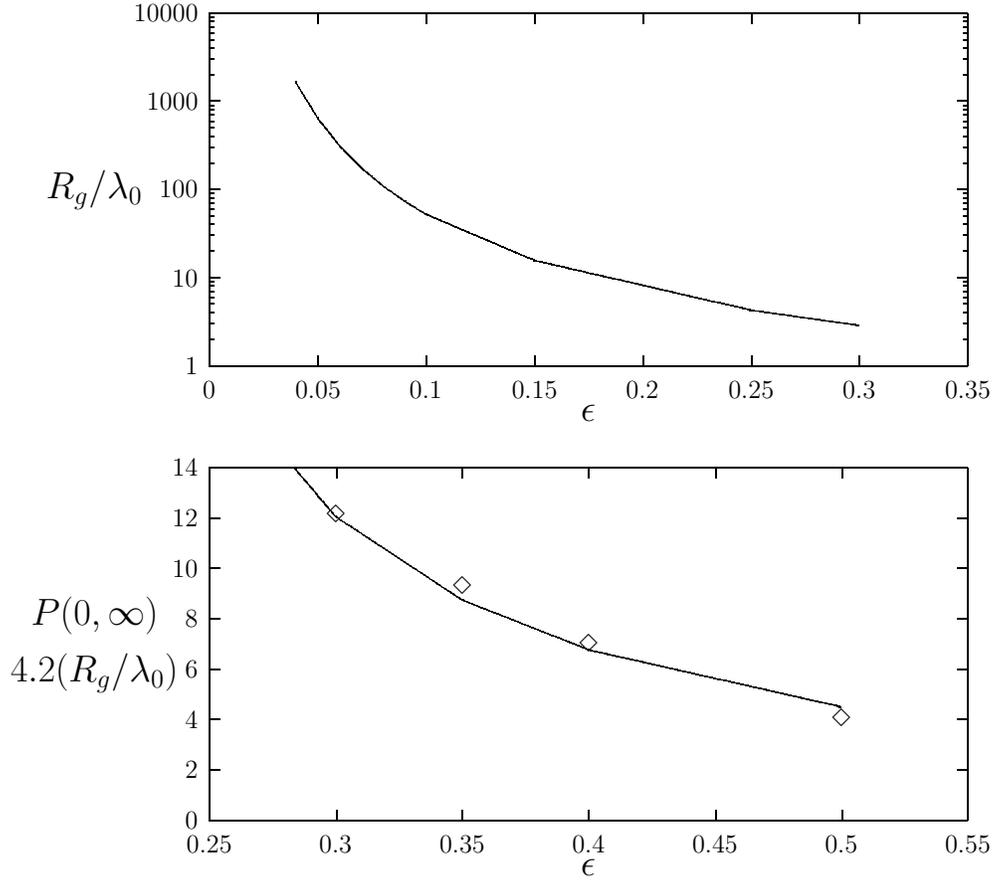,width=7in}
\vskip -8.0cm
\caption{Characteristic asymptotic grain size following a quench 
as a function of $\epsilon$. (a), estimate given by Eq. (\ref{rg}). (b),
numerical value of $P(\kappa=0,t=\infty)$ (symbols) compared also
with Eq.(\ref{rg}) multiplied by one fitted scale factor (solid line).}
\label{figRg}
\end{figure}

To further quantify these observations we have computed the
probability distribution function of stripe curvatures $P(\kappa,t)$.
The stripe curvature is defined as $\kappa = | \nabla \cdot \hat{n} |$,
where $\hat{n}$ is the unit normal to the lines of constant $\psi$.
The curvature $\kappa$ is a slowly varying quantity away from defect
cores, and only these regions are used to compute $P(\kappa,t)$ by the
filtering method described in ref. \cite{re:boyer01b}
Figure \ref{figcurv} shows our results for $\epsilon = 0.4$ and
$\epsilon = 0.5$. In both cases the distribution converges at long times
towards a limiting
curve of finite width, thus indicating that asymptotic configurations contain
many curved stripes and are disordered at large scales, or "glassy".
This behavior is to be contrasted with that of
a coarsening system in which $P(\kappa, t\rightarrow \infty)$ would approach
a delta function at $\kappa = 0$. We take $P(\kappa = 0,t \rightarrow
\infty)$ as a measure of the linear scale of the structure or typical domain
size and compare its value with the pinning radius $R_{g}$ given in
Eq. (\ref{rg}). Figure \ref{figRg}b shows the numerical results
together with $R_g$ multiplied by a (fitted) scale factor approximately equal
to $4$. The pinning radius $R_g$ increases extremely quickly with decreasing
$\epsilon$, in agreement with the numerical calculations for the range
of $\epsilon$ which we can study
(computational constraints on system sizes have prevented
us from investigating the region $\epsilon < 0.30$).
We have checked that the glassy configurations at long times do not
result from numerical pinning; the results are not modified when the
grid spacing is halved to $\Delta x = \lambda_{0}/16$.

Although other types of defects (e.g., dislocations and +1/2
disclinations) may also become pinned,
and thus contribute to the overall stability
of glassy configurations, the predominance of grain boundaries over
other defects seems to be a generic feature of the Swift-Hohenberg model
(see Figures \ref{figheating}a,b and ref. \cite{re:boyer01b}).
Furthermore it is likely that a similar dependence between the speed
of the defect and $\epsilon$ will
hold for the motion of other topological defects (except for dislocation
climb). Hence we argue that
a defected configuration of stripes does not
macroscopically order following a quench to a finite value of
$\epsilon$. Asymptotic long time configurations appear to exhibit a
labyrinthic and partially
disordered structure with many immobile defects that do not anneal away.
These disordered configurations resemble those of a structural glass at zero
temperature which lack long range order (translational or
orientational) order. They become spontaneously trapped in metastable
configurations that are very different from the configuration of lowest
free energy (all stripes parallel to each other, or a "crystalline" state).

Not all grain boundaries in a glassy configuration are $90^{\circ}$
boundaries. However, we expect that grain boundaries with a
different orientation
would be pinned less efficiently ({\it i.e.}
would have a higher value of $|\alpha|$ in Eq. (\ref{estimp})). The reason is
that their stationary planar profile is smoother than that of a $90^{\circ}$
grain boundary, and therefore non-adiabatic effects are expected to
be weaker.

We finally mention that if both $\epsilon$
and $\kappa \lambda_0$ are not small compared to one, both adiabatic and
non-adiabatic terms will contain higher order analytic corrections which we
have not calculated.

\section{Motion at finite temperature}
\label{sec:finite_temp}

Given the results of Section \ref{sec:pinning}, it is natural to study the
effect of random fluctuations added to Eq. (\ref{sh}).
Small amplitude fluctuations will allow activated motion of grain boundaries,
and in general unpinning. We consider in this section the stochastic
Swift-Hohenberg model
\begin{equation}
\label{shnoise}
\frac{\partial \psi}{\partial t} =
\epsilon\psi-\frac{1}{k_0^4} (k_0^2+\nabla^2)^2\psi-\psi^3\ +\eta(\vec{r},t)\ ,
\end{equation}
where $\eta$ is a Gaussian and white random noise of zero mean and variance
\begin{equation}\label{gauss}
\langle\eta(\vec{r},t)\eta(\vec{r}^{~ \prime},t')\rangle =
2F\delta(\vec{r}-\vec{r}^{~ \prime})
\delta(t-t').
\end{equation}
The noise intensity $F$ is proportional to the (dimensionless)
temperature according to the fluctuation-dissipation theorem.
In what follows, $F$ and $\epsilon$ are considered as
independent parameters, although they might be related in some particular
physical systems. The stochastic Swift-Hohenberg model
has been used to study hydrodynamic fluctuations near
onset of Rayleigh-B\'enard convection \cite{re:ahlers81}, and
thermal fluctuations of molecular origin in
lamellar phases of diblock copolymers \cite{re:hohenberg95}.

The stationary states of Eq. (\ref{shnoise}) in two spatial dimensions
have been studied in refs. \cite{re:toner81,re:elder92b}.
Above a critical noise intensity $F_c$ (that depends on $\epsilon$),
the system is disordered (lacks both translational and orientational
long ranged order). Below $F_c$
a stripe phase with long ranged orientational order but no
translational order was found. Only at $F=0$ the system was seen to exhibit
both translational and orientation long ranged order.
In what follows we focus on defect dynamics in the range
$0< F \ll F_c$, so that the local stripe pattern is not very distorted.

We first derive the stochastic amplitude equations for
a $90^{\circ}$ grain boundary. Following Graham \cite{re:graham74}, we
approximate the effect of the noise on the amplitudes by
projecting it along the two slow modes of the deterministic equation
and neglecting any contribution
arising from couplings and resonances between noise and
fast variables \cite{re:drolet98,re:drolet01}.
We start by writing the random function as,
\begin{eqnarray}
\eta(\vec{x},t)=&\frac{1}{2}&\left[e^{ik_0x}\ \tilde{\eta}_A(X_A,Y_A,T)
\right.\nonumber\\
&+&\left. e^{ik_0y}\ \tilde{\eta}_B(X_B,Y_B,T)\ +\ {\rm c.c.}\right]\ ,
\end{eqnarray}
where the slow variables $X(Y)_{A,B}$ are given by Eq. (\ref{slow}), and
$\tilde{\eta}_A$ and $\tilde{\eta}_B$ are two independent complex
random processes that satisfy the relations,
\begin{eqnarray}
&\langle&\tilde{\eta}_A\rangle=\langle\tilde{\eta}_B\rangle=0,\
\langle\tilde{\eta}_A^2\rangle=\langle\tilde{\eta}_A\tilde{\eta}_B\rangle=
\langle\tilde{\eta}_A\tilde{\eta}_B^*\rangle=0, \nonumber\\
&\langle&\tilde{\eta}_A\tilde{\eta}_A^*\rangle=
\langle\tilde{\eta}_B\tilde{\eta}_B^*\rangle=
2F\delta(\vec{x}-\vec{x}^{~ \prime})\delta(t-t').\nonumber
\end{eqnarray}
It is implicit in the decomposition that $F$ is small enough so
that well defined stripes exist locally. On the other hand,
$F$ has to be large enough so that $\tilde{\eta}_A$ and $\tilde{\eta}_B$ are
not negligible in the solvability conditions at order
$\epsilon^{3/2}$ \cite{fo:db3_1}. Given both assumptions,
Equations (\ref{ampa}) and (\ref{ampb}) straightforwardly generalize to
\begin{eqnarray}
\label{ampan}
\frac{\partial A}{\partial t}=&-&\frac{\delta F_{gb}}{\delta \bar{A}}\nonumber\\
&-&\frac{1}{4\lambda_0^2} \int_{x}^{x+\lambda_0}{dx'}
\int_{y}^{y+\lambda_0}{dy'}\ A^3(x',y',t)\ e^{i2k_0x'} \nonumber\\
&+& \tilde{\eta}_A,
\end{eqnarray}
\begin{eqnarray}
\label{ampbn}
\frac{\partial B}{\partial t}&=&-\frac{\delta F_{gb}}{\delta \bar{B}}\nonumber\\
&-&\frac{3}{4\lambda_0^2} \int_{x}^{x+\lambda_0}{dx'}
\int_{y}^{y+\lambda_0}{dy'}[A^2B e^{i2k_0x'}+\bar{A}^2B
e^{-i2k_0x'}]\nonumber\\ 
&+& \tilde{\eta}_B.
\end{eqnarray}

We can now estimate the escape rate of a grain boundary over the potential
barrier of Eq. (\ref{motionlaw}). In order to do so, we need to
estimate the projection of the noise intensity in Eqs.
(\ref{ampan}) and (\ref{ampbn}) on the coordinate $x_{gb}(t)$ implicitly
defined by Eqs. (\ref{sola}) and (\ref{solb}). A rough estimate that is
sufficient for our purposes can be obtained by using Eqs. (\ref{sola}) and
(\ref{solb}) as the trial solution of Eqs. (\ref{ampan}) and (\ref{ampbn}).
Focusing on $x_{gb}$ alone ignores possible boundary broadening because
of fluctuations, or roughening. Both phenomena will be important for grain
boundary motion above the pinning point, but their contribution is probably
less important in the immediate vicinity of the pinning transition. By
substituting Eqs. (\ref{sola}) and (\ref{solb}) into Eqs. (\ref{ampan}) and
(\ref{ampbn}), we find
\begin{equation}
\dot{x}_{gb}=\frac{\epsilon}{3k_0^2D(\epsilon)}
\ \kappa^2\  - \frac{p(\epsilon)}{D(\epsilon)}\cos(2k_0 x_{gb}+\phi)
+\tilde{\eta}\ ,
\label{noisymotionlaw}
\end{equation}
with $\tilde{\eta}$ a (real) random white Gaussian noise satisfying,
\begin{equation}\label{effectivenoise}
\langle\tilde{\eta}\rangle=0,\quad
\langle\tilde{\eta}(t)\tilde{\eta}(t')\rangle=
2F'\ \delta(t-t'),\quad F'=F/[2D(\epsilon)R_{gb}]
\end{equation}
where $R_{gb}$ is the grain boundary perimeter. As expected, the
intensity of the fluctuations on the global coordinate $x_{gb}$ is
proportional to $1/R_{gb}$. Equation
(\ref{noisymotionlaw}) is a straightforward generalization of
Eq. (\ref{motionlaw}), and is formally analogous to the equation that
describes the one dimensional motion of a Brownian particle in a periodic
potential of amplitude $2p(\epsilon)/[2D(\epsilon)k_0]$.

Equations (\ref{noisymotionlaw}) and (\ref{effectivenoise}) can be recast as
\begin{eqnarray}
\dot{x}_{gb}&=&\left(\frac{k_0 F_0}{2D}\right) R_g \kappa^2 
- \left(\frac{k_0 F_0}{2D}\right)\frac{1}{R_g}\cos(2k_0 x_{gb}+\phi)\nonumber\\
&\ &+\frac{1}{\sqrt{2D}}\left(\frac{F}{R_{gb}}\right)^{1/2}\xi\ .
\label{noisymotionlaw2}
\end{eqnarray}
The random term $\xi$ is such that
$\langle\xi\rangle=0$ and $\langle\xi(t)\xi(t')\rangle=2\delta(t-t')$.
We have also used Eq. (\ref{rg}) to eliminate $p(\epsilon)$ from
Eq. (\ref{noisymotionlaw}), and we have defined
\begin{equation}
\label{F0}
F_0=\frac{2\epsilon}{3k_0^3R_g}.
\end{equation}
Consider the situation where grain boundaries are pinned at $F=0$.
Since $\kappa<\kappa_g$, the first term of the right-hand-side of
Eq. (\ref{noisymotionlaw2}) is not dominant and
the potential barrier that a pinned defect
of size $R_{gb}$ has to overcome is of the order of $F_0/R_g$.
The stochastic problem
is now an escape problem over this potential barrier given the intensity
of the noise term in Eq. (\ref{noisymotionlaw2}). The Kramers rate of escape 
is given by
\begin{equation}
\label{kramers}
r\sim\exp\left(-\frac{F_0}{F}\frac{R_{gb}}{R_g}\right).
\end{equation}
Therefore a noise intensity
\begin{equation}
F=R_{gb}\ \frac{F_{0}}{R_g} \sim R_{gb}\ k_0^{-1}\epsilon^2\
e^{-|\alpha|/\sqrt{\epsilon}}
\end{equation}
is required to un-pin a grain boundary of length $R_{gb}$.

\section{Slow coarsening dynamics: Dependence on temperature and quench depth.}
\label{sec:coarsening}

We use here the results of Sections \ref{sec:pinning} and
\ref{sec:finite_temp} to provide a possible interpretation of conflicting
results concerning domain coarsening of stripe phases.
We recently studied this issue by numerically solving the {\it noiseless}
Swift-Hohenberg equation (Eq. (\ref{sh})) {\it in the limit} 
$\epsilon\rightarrow0$ \cite{re:boyer01b}. 
Our numerical results suggested that the characteristic
scale of the structure (or the typical size of ordered domains) increases 
as $t^{1/z}$, with $z=3$. That value of the exponent
was interpreted to follow from the dominant motion of grain boundaries
through a background of curved stripes. In disordered configurations,
the curvature of stripes is set by a distribution of largely immobile
+1/2 disclinations. According to Eq. (\ref{motionlaw}), the motion of
grain boundaries is driven by stripe curvature, and acts to reduce the overall
curvature by replacing regions of curved stripes by straight ones of
a different orientation. It also reduces the
disclination density whenever their core region is swept by a moving grain
boundary. In the limit $\epsilon \ll 1$ we computed several measures of
the linear scale, including moments of $P(\kappa,t)$, moments of the
structure factor of the order parameter, and the average distance between
defects. They were all found to become
proportional to each other, and to grow as a power law of time
with an exponent $1/3$.

Grain boundary motion as described in Section \ref{sec:pinning} was used
to provide an interpretation for the value $z=3$.
Since +1/2 disclinations generate roughly axisymmetric patterns of stripes
around them, the
characteristic stripe curvature in any given configuration
is proportional to the
inverse characteristic distance between disclinations.
Under the self-similarity hypothesis, the distance between disclinations is
proportional to $l(t)$, hence $\kappa \sim 1/l(t)$.
If grain boundaries are the class of defect the motion of which controls
asymptotic coarsening,
then the coarsening exponent can be inferred by dimensional analysis
of Eq. (\ref{motionlaw}).
In the limit $\epsilon \rightarrow 0$, the oscillatory
term in the right hand side of Eq. (\ref{motionlaw}) can be neglected and
we simply have $dl/dt \propto l^{-2}$ or $l(t) \sim t^{1/3}$, in
agreement with the numerical solution of Eq. (\ref{sh}).

Equation (\ref{motionlaw}) shows that this result
changes qualitatively further from onset.  As $\epsilon$ increases
the pinning potential energy barrier $p(\epsilon)$
increases extremely fast, and important corrections to scaling are to be
expected. For finite $\epsilon$ and short times many defects are present,
therefore the characteristic curvature of the stripes
is very large, and the first term in the right hand side of
Eq. (\ref{motionlaw}) dominates. As coarsening proceeds,
the characteristic curvature decreases until it reaches the critical
value $\kappa_g$ given by Eq. (\ref{rg}). At that point
the typical velocity of a grain boundary vanishes, although the
system is still disordered. Therefore one would expect that coarsening would
stop when $l(t)$ is of the order of $R_{g}$. This is precisely the results
shown in Fig. \ref{figRg} with only one adjustable parameter (a scale
factor relating $R_{g}$ given by Eq. (\ref{rg}) to $l(t)$ determined
numerically from the distribution of stripe curvatures).

When random fluctuations are considered
($F > 0$ in Eqs. (\ref{shnoise})-(\ref{gauss})),
some of the grain boundaries in a frozen configuration are expected to resume
motion. We argue that the structure will continue coarsening until the
average domain size reaches a new characteristic size $l_F > R_g$
that can be estimated as follows.
We write a general phenomenological evolution equation for the domain size
$l(t)$ directly from Eq. (\ref{noisymotionlaw2}):
\begin{eqnarray}
\label{eqR}
\frac{d l}{dt}&=&\left(\frac{k_0F_0}{2D}\right)\frac{R_g}{l^2}
-\left(\frac{k_0F_0}{2D}\right)\frac{1}{R_g}\cos(2k_{0} l +\phi)\nonumber\\
&\ &+\frac{1}{\sqrt{2D}}\left(\frac{F}{l}\right)^{1/2}\xi \ ,
\end{eqnarray}
where we have assumed that, prior to pinning, the various length scales remain
approximately proportional to each other. Recall from
Eq. (\ref{kramers}) that $F=F_{0}$ is required to unpin a
configuration obtained in the absence of noise,
for which $l(t) \sim  R_{g}$.
According to Eq. (\ref{eqR}), coarsening proceeds if $F > F_{0}$ until
a new characteristic pinning size is reached given by  $F_0 l_{F}/(FR_g)=1$ or
\begin{equation}
\label{R1}
l_{F} = R_{g}\ \frac{F}{F_0} \sim k_{0} F
\frac{e^{|\alpha|/\sqrt{\epsilon}}}{\epsilon^2}.
\end{equation}
After reaching the scale $l_F$, domains are expected to coarsen very
slowly by thermal activation. When a grain boundary overcomes one pinning
barrier, the linear extent of the corresponding domain typically increases
by an amount of order $\lambda_0/2$. Hence $dl/dt\sim \lambda_0 r$, where
$r$ is given by Eq. (\ref{kramers}) with $R_{gb}$ replaced by $l$. Hence
domains are expected to grow logarithmically in time according to
\begin{equation}
l(t) \sim F\ln (t/F)\quad {\rm for}\ l \gg l_F.
\end{equation}

A numerical solution of Eq. (\ref{shnoise}) yields results qualitatively
consistent with
those presented above. Figure \ref{figheating}a shows a
configuration of the order parameter field $\psi$ obtained for
$F=0$ and $\epsilon=0.4$ starting from random initial conditions.
The configuration shown corresponds to very late times $t = 2.3\times 10^5$ at
which point
all defects are practically immobile, and domain growth has stopped.
We then set $F = 0.00318$, and the integration is
continued. The order parameter configuration
$t=10^5$ time units later is shown in Fig. \ref{figheating}b. The average
domain size has increased substantially. Many
grain boundaries have a $90^{\circ}$ orientation (like in Fig.
\ref{figheating}a), and roughening is limited or nonexistent.
We have determined the average domain size $l$ from the probability
distribution function of the quantity 
$\zeta=\psi^2+(\vec{\nabla}\psi)^2/k_0^2$.
Figure \ref{figappend.b} shows the probability distribution function
corresponding to a perfectly
ordered configuration, as well as to partially disordered configurations.
The inverse linear scale $1/l$, proportional to the defect
density $\rho_d$, is extracted from the difference between these curves, 
as detailed in Appendix \ref{ap:b}. 
As shown in Fig. \ref{figRF} domain growth is very slow, possibly
logarithmic, although a precise check of this behavior is problematic.

\begin{figure} 
\centerline{\epsffile{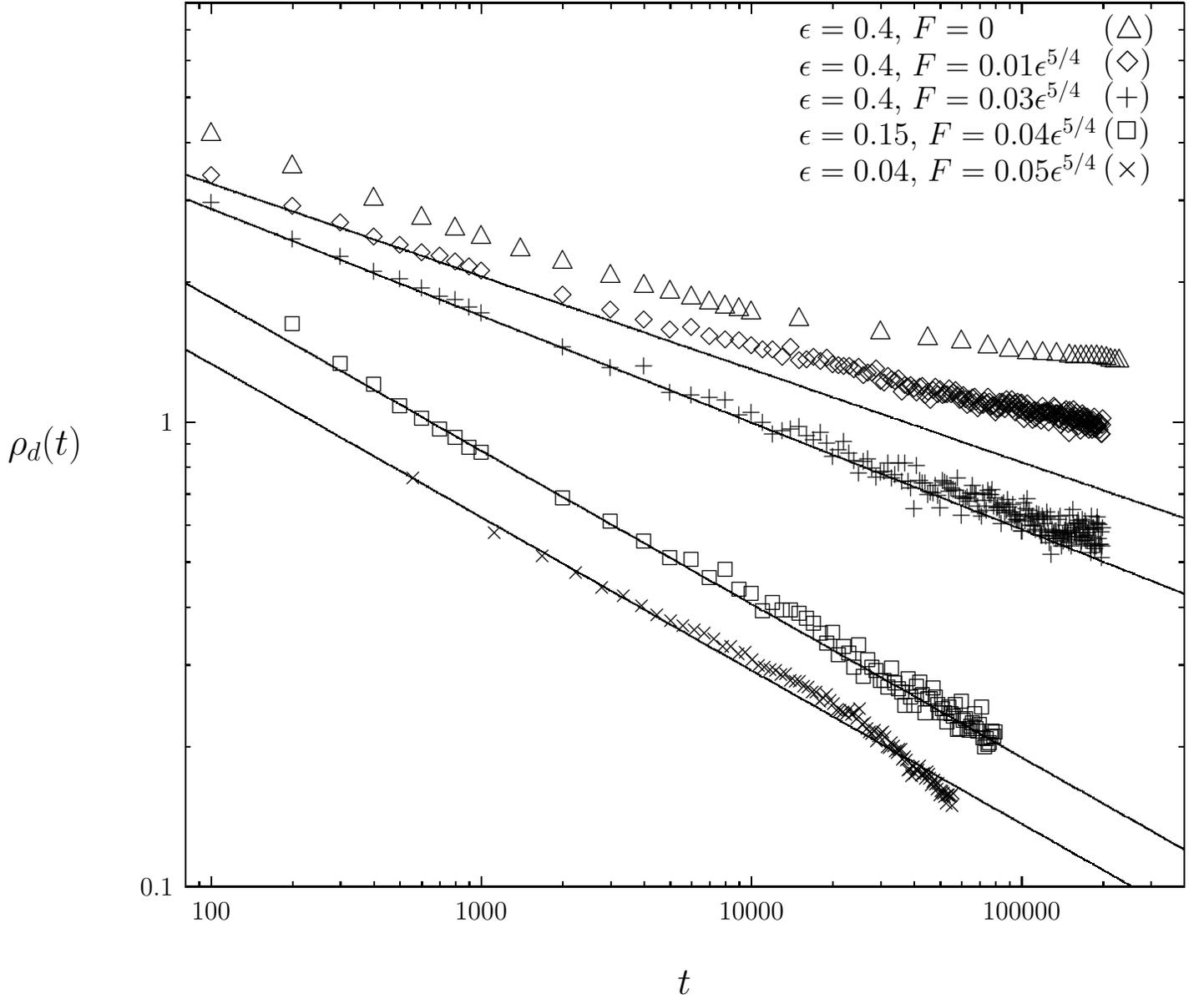}}
\caption{Defect density in arbitrary units as a function of time
for several values of $\epsilon$ and $F$. 
The straight solid lines are guides to the eye with slopes, from bottom 
to top, $-0.33$, $-0.33$, $-0.23$ and $-0.20$.}
\label{figrhod}
\end{figure}

Figure \ref{figrhod} displays the evolution of the defect density $\rho_d(t)$ 
as a function of time, starting from random initial configurations.
For reference we also show the case $F = 0$. Increasing the value of
$\epsilon$ leads to smaller effective exponents, whereas increasing $F$
has the opposite effect. For sufficiently small $\epsilon$, we find
$z = 3$ independent of the value of $F$.
The two bottom curves correspond to systems that are
close enough to onset, and hence either $R_{g}$ or $l_{F}$ is very large
compared with the linear size of the system.
We show our results for $\epsilon=0.04$ (averaged over 40 independent runs) 
and for $\epsilon=0.15$ (averaged over 15 independent runs). The solid
line closest to these two curves has a slope of $-1/3$.
The downward deviation from linearity at long times at $\epsilon=0.04$ is 
a typical manifestation of finite size effects
(this long time behavior and its dependence on the system size
was studied in detail in ref. \cite{re:boyer01b}).

With increasing $\epsilon$ and/or decreasing $F$, pinning becomes more
pronounced as evidenced the lower effective slopes of
the three upper curves in Fig. \ref{figrhod} (the results
are averages over six independent runs, each curve corresponding to the
same value $\epsilon=0.4$). The top curve corresponds to a system without
fluctuations for which $\rho_d$ was computed with the method described
in ref. \cite{re:boyer01b}.
The density starts decaying roughly as an inverse power law, with an 
effective exponent much smaller than $-1/3$ (the top solid line
has a slope of $-1/5$), and after a crossover saturates at long times 
indicating pinning. When small amplitude noise is added (curve below 
denoted by diamonds), the initial behavior is similar to that of $F=0$, 
and the decay rate also slows down considerably at long times (where we
would predict logarithmic growth, {\it i.e.} $\rho_d\sim 1/\ln (t)$). 
The curve below, denoted by plus signs, corresponds to a noise intensity 
three times larger than the previous case. Its initial decay is
slightly faster (it can be fitted with
an effective exponent  $1/z_{eff} \simeq -0.23$ as shown with the solid
line in the figure), and the upwards deviations at long times are less
pronounced. This behavior is in qualitative agreement with the the 
expectation that defects overcome pinning barriers more readily at higher 
noise intensities and pinning is postponed to longer times when $l\sim l_F$.
However, the effective initial decay is slower than $t^{-1/3}$ which we 
interpret as a crossover effect resulting from non-adiabaticity.

In summary, our results for large $\epsilon$ are in agreement with earlier
numerical results performed at $\epsilon = 0.25$, showing that coarsening laws
are very slow and depend on the presence of thermal fluctuations. 
We argue here that a coarsening exponent can be properly determined only 
in the small $\epsilon$ limit, where the phase ordering kinetics 
is self-similar. Our results support that the exponent $z = 3$ is
independent of $F$ for sufficiently small $\epsilon$, when pinning effects
are negligible ($R_{g}$ much larger than the linear size of the system).

\section{Conclusions}
\label{sec:conclusions}

We have shown that the Swift-Hohenberg model of Rayleigh-B\'enard
convection exhibits glassy properties in spatially extended systems.
In the absence of fluctuations, and following a parameter quench across
threshold, random initial configurations do not evolve into completely
ordered states, a single plane-wave or crystalline state. Instead, they
reach disordered metastable configurations in which topological defects, 
mainly grain boundaries and disclinations, fail to annihilate and remain 
with finite density. It appears that the formation
of these glassy configurations in a quenched disorder-free system can 
be accounted for by the finite separation between
\lq\lq fast" length scales of the structure (associated with stripe
periodicity), and \lq\lq slow" scales (associated with the extent of defect
envelopes). Since at a finite distance from threshold the ratio between 
these two scales is finite, non-adiabatic
effects lead systematically to defect pinning in an infinite system. 
Fluctuations allow unpinning and a certain amount of
\lq\lq crystallization", albeit through an asymptotically slow
activated motion of grain boundaries and other defects. 

The present framework is far too simple to be used in the prediction of a
glass transition temperature, if such a transition exists. In some respects,
the situation
just described is instead very similar to that of domain growth
in random fields in dimension larger than two
\cite{re:natterman88,re:bouchaud98}.
There, domain walls separating magnetized domains are pinned by
fixed impurities and evolve by thermal activation to other more favorable
configurations. The phenomenological pinning energy of a domain of size
$R$ grows as $\Upsilon R^{\theta}$, were $\theta$ depends on the
problem considered, yielding an escape rate given by
$r\sim\exp(-\Upsilon R^{\theta}/k_BT)$, equation that is formally
analogous to Eq. (\ref{kramers}) with $\theta=1$. 
Two crucial differences are that our system is glassy even in two
dimensions, and that defects do not need any disorder to become pinned.

The consequences of defect pinning on the intermediate time regime 
corresponding to domain coarsening have also been
investigated. A universal coarsening
exponent can be determined close to threshold only, where we obtain $z=3$.
Coarsening stops when the linear size of the system is
larger than the characteristic domain size for pinning. In this situation,
an intermediate crossover regime is anticipated with lower effective
coarsening exponents, as is observed in numerical solutions of the model.
Crossover effects induced by pinning can be reduced by either increasing 
the intensity of the fluctuations or approaching threshold. 

We note that some of
our conclusions as well as our interpretation of the numerical results
are based on the analysis of a particular type of defect,
namely a grain boundary separating two
domains with differently oriented stripes. We think it is likely that
similar non-adiabatic corrections to defect motion will appear for
dislocation glide or disclination motion, leading to similar non perturbative
corrections in $\epsilon$ to the speed of the defect.

We believe more generally that pinning through non-adiabatic effects is 
likely to be a feature of a wide variety of pattern forming systems, 
and is not limited to the particular model treated here. 
Block copolymer melts, for instance, provide an interesting case in which 
the results obtained could have practical implications (see Appendix A for a 
summary of the relevant equations and their relationship with the model 
studied here).
We also mention here that results qualitatively similar to ours have been 
reported for a model with competing 
interactions (describing ferromagnetic films), that is defined by the equations 
of Appendix A with a different form of the Green's function $G$ 
\cite{re:sagui94}.
There, frozen polycrystalline configurations of stripe patterns
were observed for deep quenches as well, whereas the system could reach
an ordered state for shallow quenches. This same model was also able to
predict the formation of a frozen phase composed of polydisperse 
droplets with a near-hexagonal arrangement
\cite{re:sagui95}, as previously observed in experiments on a Langmuir 
monolayer \cite{re:seul94}. However, the pinning mechanism involved in 
this last case is probably different than the one discussed in the present
paper since the patterns are no longer locally periodic. 
Nevertheless, we would expect that our main conclusions can
be readily extended to other systems with periodic structures 
such as hexagonal patterns \cite{re:sagui95,re:elder01}.

\begin{acknowledgments}
This research has been supported by the U.S. Department of Energy, contract
No. DE-FG05-95ER14566.
\end{acknowledgments}

\appendix 
 
\section{Mean-field model of a symmetric block copolymer melt}
\label{ap:a}

We briefly recall in this appendix known results about the relationship
between the mean field description of a block copolymer melt, and the 
amplitude equation for Swift-Hohenberg model (Eq. (\ref{sh})) at first order 
in $\epsilon$. The dynamics of micro-phase separation of block copolymers
is often modeled by a time-dependent Ginzburg-Landau equation for a 
conserved order parameter \cite{re:leibler80,re:ohta86},
\begin{equation}\label{cahnhil}
\frac{\partial\psi(\vec{r},t)}{\partial t}=M\nabla^2
\frac{\delta F}{\delta\psi(\vec{r},t)},
\end{equation}
where
\begin{eqnarray}
F&=&\int d\vec{r}\ \left(-\frac{r}{2}\psi^2+\frac{u}{4}\psi^4+\frac{K}{2}
(\nabla \psi)^2\right)\nonumber\\
&\ &+\frac{B}{2}\int\int d\vec{r}d\vec{r}^{~ \prime} 
\psi(\vec{r},t)G(\vec{r},\vec{r}^{~ \prime}) \psi(\vec{r}^{~ \prime},t).
\end{eqnarray}
$G$ is the Green's function of the Laplacian operator
$\nabla^2 G(\vec{r},\vec{r}^{~ \prime})
=-\delta(\vec{r}-\vec{r}^{~ \prime})$
and $M$ a constant mobility or Onsager coefficient. The scalar order 
parameter $\psi$ is the local 
monomer concentration difference between the two chemical species.
Following Ref. \cite{re:shiwa97} we set $r = 2+\epsilon,
u = 1, K = 1/k_{0}^{2}$ and $B = k_{0}^{2}$.
Two independent parameters $k_0$ and (small) $\epsilon$ remain. 
Equation (\ref{cahnhil}) reduces to
\begin{equation}
\label{tdgl2}
\frac{1}{M}\frac{\partial\psi}{\partial t}=\nabla^2[-(2+\epsilon)\psi+\psi^3
-\frac{1}{k_0^2}\nabla^{2} \psi]-k_0^2 \left( \psi - \psi_{\infty} \right),
\end{equation}
where $\psi_{\infty}$ is the boundary condition at infinity. In
most studies, it is customary to set $\psi_{\infty} = 
\langle \psi \rangle$, the spatial average of $\psi$ over the sample.
We introduce the amplitude $A$ of slightly modulated waves through,
\begin{equation}
\psi(\vec{r},t)=\frac{1}{2}[A(\vec{r},t)\ e^{ik_0x}+\ {\rm c.c.}].
\end{equation}
A multi-scale analysis of Eq. (\ref{tdgl2}) in the limit $\epsilon \ll 1$ was 
conducted by Shiwa \cite{re:shiwa97}.
Setting $M=1/k_0^2$, the resulting equation for the amplitude is
\begin{equation}
\frac{\partial A}{\partial t}=\epsilon A+\frac{4}{k_0^2}
\left(\partial_x-\frac{i}{2k_0}\partial_y^2\right)^2A
-\frac{3}{4}|A|^2A\ ,
\end{equation}
which is identical to the amplitude equation of the Swift-Hohenberg 
model. Note that the only effect of the conservation law on the local
part of the free energy (the Laplacian operator in front of the square 
bracket in Eq. (\ref{tdgl2})) is a renormalization of the mobility $M$.
The quantities $\epsilon$ and $k_{0}$ defined above play the same role as 
the same coefficients in the Swift-Hohenberg model ({\it i.e.} the 
dimensionless distance to threshold and the dominant wavenumber of the 
structure, respectively.)

\section{Calculation of the defect density in the presence of fluctuations}
\label{ap:b}

\begin{figure}
\epsfig{figure=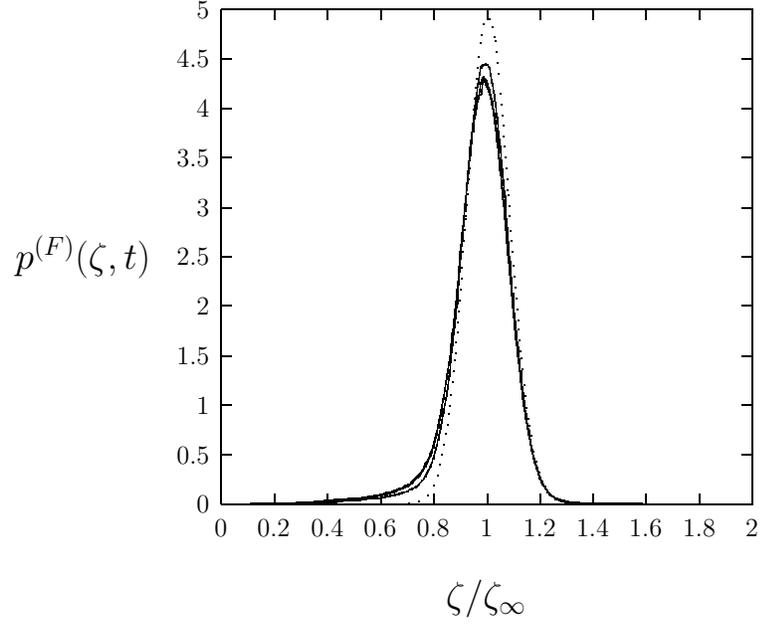,width=7in}
\vskip -3cm
\caption{Probability distribution function of $\zeta$ defined
by Eq. (\ref{zeta}) with $\psi$ the solution of Eq. (\ref{shnoise}), 
for $F=0.00636$ and $\epsilon=0.4$. The dotted line corresponds to a
single plane wave with superimposed fluctuations,
while the solid line correspond to disordered 
configurations obtained from random initial conditions 
(at times $t=5 \times 10^3$, $10^4$ and $10^5$, respectively).}
\label{figappend.b}
\end{figure}

\begin{figure}
\epsfig{figure=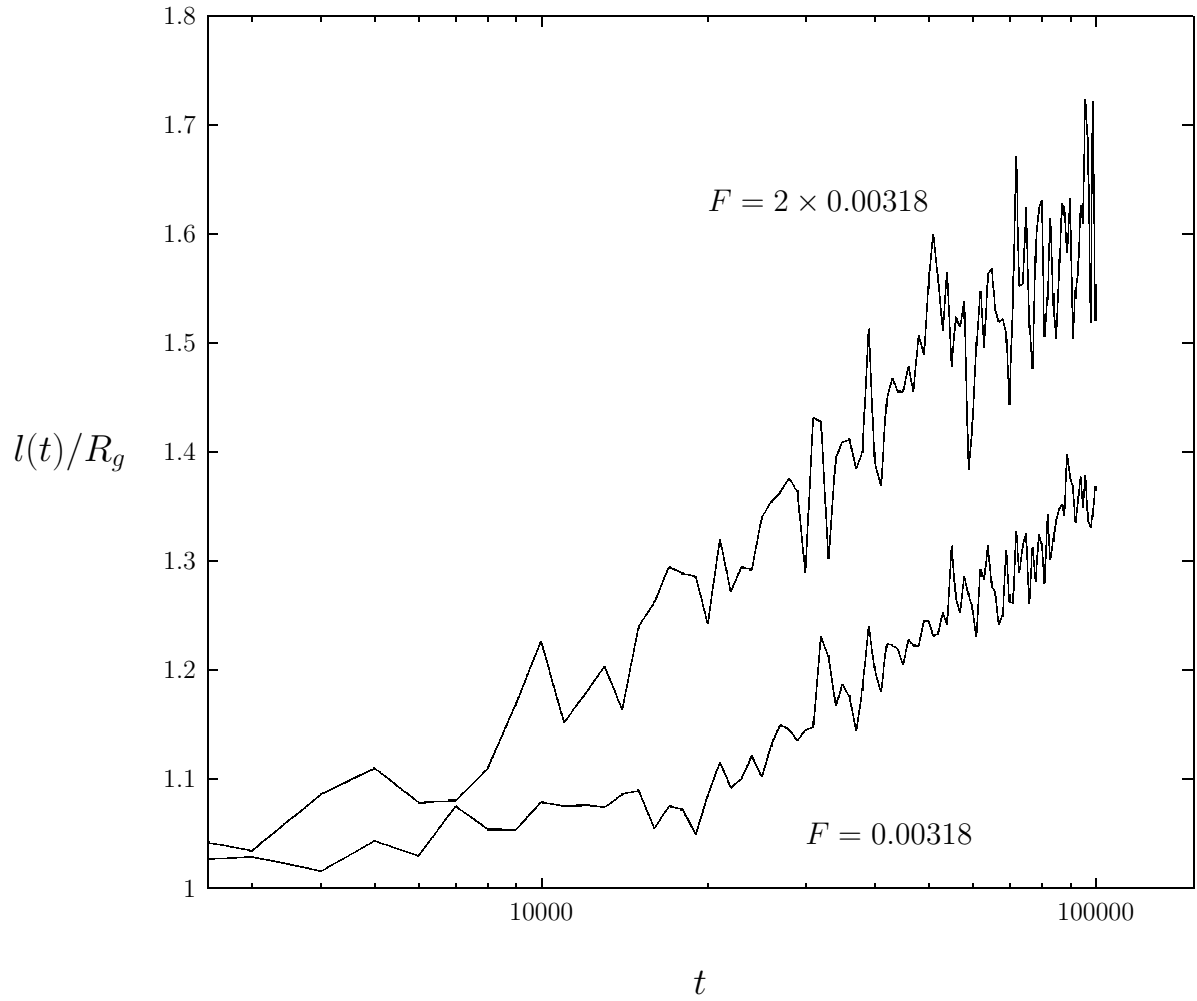,width=7in}
\vskip -6cm
\caption{Characteristic domain size as a function of time 
at $\epsilon=0.4$ and $F \neq 0$ as indicated in the figure. The initial 
condition at time $t=0$ is a glassy 
configuration obtained from a previous run with $F = 0$.}
\label{figRF}
\end{figure}

Computation of the domain size from
the probability distribution of stripe curvature is delicate in the presence
of noise. We have used a different method than that used for $F=0$.
We introduce an effective squared amplitude $\zeta(\vec{r},t)$ by
\begin{equation}
\label{zeta}
\zeta=\psi^2+(\vec{\nabla}\psi)^2/k_0^2\ .
\end{equation}
For a perfectly ordered system consisting of a plane-wave solution 
of the Swift-Hohenberg equation and $F=0$, the probability distribution 
function of $\zeta$ is a delta function at $\zeta_{\infty}=4\epsilon/3$.
When $F>0$ the probability distribution function of $\zeta$ even for a plane 
wave $p_{\infty}^{(F)}(\zeta)$ is broader because of \lq\lq phonon'' 
excitations. The function $p_{\infty}^{(F)}$ is plotted in 
Figure \ref{figappend.b} (with dotted lines),
and is then used as a reference curve for a fixed $F$. 
In a partially disordered configuration, the presence of defects and curved 
stripes further broadens the probability distribution function to
$p^{(F)}(\zeta,t)$ (solid lines of Figure \ref{figappend.b}). The difference 
between the two curves is related to the degree or disorder beyond
small fluctuations away from a perfectly ordered structure. We define 
the defect density by $\rho_d(t)={\rm Max}\{p_{\infty}^{(F)}(\zeta),\ 
\zeta\} - {\rm Max}\{p^{(F)}(\zeta,t),\ \zeta\}$. 
Since grain boundaries 
are seen to be the major contribution to defect density, one can introduce a
characteristic length scale $\rho_d^{-1}$, which is further identified with
the characteristic size, or domain size $l$. 
In Figure \ref{figRF}, $l$ has been normalized so that $l(t=0)=R_g$
(at the beginning of the heating process) where $R_g$ is 
computed from the probability distribution function of stripe curvatures 
at $F=0$.

\bibliographystyle{prsty}

\begin{thebibliography}{10}
\bibitem{fo:db3_1}
$F$ can be estimated to be of the order $F=u\epsilon^{5/4}$, with $u\le 1$. See
  also Ref. \protect\cite{re:graham74}.
\end{thebibliography}

\newpage
\onecolumn

\newpage

\newpage

\newpage

\newpage

\newpage

\end{document}